\documentclass[draft]{agujournal2019}

\usepackage{multirow}
\usepackage{url}
\usepackage{soul}
\draftfalse
\journalname{Earth and Space Science}

\begin{document}
\title{Comparison of automated crater catalogs for Mars from \citeA{benedixDerivingSurfaceAges2020} and \citeA{leeAutomatedCraterDetection2021}}
\authors{C. Lee \affil{1}}
\affiliation{1}{University of Toronto}
\affiliation{1}{60 St. George St., Toronto, Ontario, Canada}
\correspondingauthor{Christopher Lee}{clee@atmosp.physics.utoronto.ca}
\begin{keypoints}
\item The apparent performance of automatic crater detection algorithms is sensitive to the choice of metric used to gauge the performance.
\item Feature Detection Algorithms trained to find spatial features should be used with appropriately projected images where possible.
\item As neural networks extend crater detection beyond practical human limits, validating their results with independent networks is critical.
\end{keypoints}
\begin{abstract}
Crater mapping using neural networks and other automated methods has increased recently with automated Crater Detection Algorithms (CDAs) applied to planetary bodies throughout the solar system. 
A recent publication by \citeA{benedixDerivingSurfaceAges2020} showed high performance at small scales compared to similar automated CDAs but with a net positive diameter bias in many crater candidates. I compare the publicly available catalogs from \citeA{benedixDerivingSurfaceAges2020} and \citeA{leeAutomatedCraterDetection2021} and show that the reported performance is sensitive to the metrics used to test the catalogs. I show how the more permissive comparison methods indicate a higher CDA performance by allowing worse candidate craters to match ground-truth craters. I show that the \citeA{benedixDerivingSurfaceAges2020} catalog has a substantial performance loss with increasing latitude and identify an image projection issue that might cause this loss. Finally, I suggest future applications of neural networks in generating large scientific datasets be validated using secondary networks with independent data sources or training methods.
\end{abstract}

\section*{Plain Language Summary}
I have compared two computer programs that use neural networks to identify craters on solar system bodies. The crater catalogs created by these programs have been previously compared against an existing human-made catalog to measure their performance, but each comparison used a different method making it challenging to understand the difference between the catalogs. In this paper, I use the comparison methods from these two independent papers, discuss where the catalogs agree and disagree, and suggest why they disagree. I emphasize the need for more accurate and consistent methods to validate automatic crater-mapping programs, especially when it is impractical for humans to check the craters found by automated methods.

\section{Introduction}
\label{sec:intro}

Crater mapping using neural networks and other automated methods has increased in recent years with automated Crater Detection Algorithms (CDAs) applied to planetary bodies throughout the solar system including Mercury and the Moon \cite<e.g.,>[]{Silburt2019} Mars and Pluto \cite<e.g.,>[]{leeAutomatedCraterDetection2021,Lee2018a,benedixDerivingSurfaceAges2020} and large asteroids \cite<e.g.,>[]{latorreTransferLearningRealtime2023}. These methods approach or exceed the performance of human experts at identifying craters from imagery and topography data and have recently produced catalogs that are too large to be validated by humans in a reasonable time \cite{lagainTharsisMantleSource2021}. In this work, I compare the results from two automatically generated catalogs for Mars and two additional catalogs described below. I calculate the performance metrics against a shared ground-truth and show how the different test criteria originally developed for these catalogs \cite{leeAutomatedCraterDetection2019,benedixDerivingSurfaceAges2020} leads to different apparent quality of each catalog.

In the next section, I describe the catalogs and discuss some of the stated limitations of the neural networks and catalogs provided. I then reproduce the comparison algorithms from \citeA{leeAutomatedCraterDetection2019} and \citeA{benedixDerivingSurfaceAges2020} and use both methods to calculate the performance of each crater catalog in this comparison. I compare the performances of these catalogs as a function of crater location and size and discuss the differences in performance. I also identify a possible calculation error in \citeA{benedixDerivingSurfaceAges2020} and suggest a correction. Finally, I give my conclusions on this comparison and suggest ways to mitigate errors and improve future catalogs.

\section{Methods}
\label{sec:methods}
\subsection{Catalogs}
\label{sec:methods_catalogs}

In this work, I compare two publicly available crater catalogs generated by neural networks, \textbf{Benedix} from \citeA{benedixDerivingSurfaceAges2020} and \textbf{DeepMars2} from \citeA{leeAutomatedCraterDetection2021}. I also compare a catalog generated as part of \citeA{leeAutomatedCraterDetection2021} (\textbf{irH1K}) but not separately published, and a new catalog based on \citeA{benedixDerivingSurfaceAges2020} (\textbf{B\_Rescaled}) that I created as part of this work to correct a proposed error in the original catalog.

The \textbf{Benedix} catalog was generated by \citeA{benedixDerivingSurfaceAges2020} from infrared images of the planet using the ``You Only Look Once'' \cite{redmonYouOnlyLook2016,redmonYOLOv3IncrementalImprovement2018} neural network, reporting 328,833 detected craters at diameters between 1km and 200km, covering the surface between $\pm$65 degrees latitude. Following \cite{benedixDerivingSurfaceAges2020}, the trained network was used to identify 90 million small craters \cite{lagainTharsisMantleSource2021}, and the same authors have sought to validate the \citeA{Robbins2012a} catalog using computer-aided tools \cite{lagainMarsCraterDatabase2021}. The \textbf{Benedix} catalog was downloaded for this work from the supplementary material of \citeA{benedixDerivingSurfaceAges2020}. 

The \textbf{DeepMars2} catalog was generated by \citeA{leeAutomatedCraterDetection2021} using digital terrain models (DTMs) of the surface and co-registered infra-red imagery using a ResUNET network \cite{Diakogiannis2019}. Two separate networks were trained using the same \citeA{Robbins2012a} catalog as the ground-truth with randomly selected swatches from across the planet that were re-projected onto an orthographic projection. In addition to the combined catalog that used DTMs and IR detections, a high-resolution IR catalog was also developed in \citeA{leeAutomatedCraterDetection2021} but truncated to 3.5km diameter and above to match the DTM limitations. In the following sections, we relax this constraint and include this infra-red imagery catalog truncated to a minimum of 1km in diameter (labeled as \textbf{irH1k}). The \textbf{DeepMars2} catalog was downloaded for this work from the associated BorealisData repository for the \citeA{leeAutomatedCraterDetection2021} paper at \citeA{leeCraterCatalogsSoftware2020}. The \textbf{irH1k} catalog was created as part of \citeA{leeAutomatedCraterDetection2021} but is made available on the BorealisData repository associated with this work at \citeA{leeReplicationDataComparison2023}.

Finally, as part of this work, a possible error was identified in the \textbf{Benedix} catalog, and a correction was applied to create the \textbf{B\_Rescaled} catalog. The correction involved rescaling the calculation of the pixel-to-geophysical size, resulting in smaller crater sizes at high latitudes in the new catalog. The \textbf{B\_Rescaled} catalog is available from the BorealisData repository associated with this work at \citeA{leeReplicationDataComparison2023}.

Newer models have been developed since \citeA{benedixDerivingSurfaceAges2020} and \citeA{leeAutomatedCraterDetection2021}, but typically examine a small area of Mars as validation \cite{yangHighResolutionFeaturePyramid2022,giannakisDeepLearningUniversal2023} and provide no publicly accessible crater catalogs. There have also been improvements in surface imagery since \citeA{Robbins2012a} with improved geo-location of images \cite{Robbins2017} and higher resolution visible imagery (\url{http://murray-lab.caltech.edu/CTX/}) from the Mars Reconnaissance Orbiter Context Camera \cite{malinContextCameraInvestigation2007}.
 
In the rest of this paper, I refer to the algorithms and catalogs from \citeA{leeAutomatedCraterDetection2021}, \citeA{benedixDerivingSurfaceAges2020}, and \citeA{Robbins2012a}. To avoid misattribution, I refer to the catalogs by their names above (\textbf{Benedix}, \textbf{DeepMars2}, \textbf{irH1K}, \textbf{B\_Rescaled}, and \textbf{Robbins}) in the text and figures, and use the acronyms L19 \cite{leeAutomatedCraterDetection2019} and B20 \cite{benedixDerivingSurfaceAges2020} to refer to the comparison methodology developed by those authors.

\subsection{Ground Truth Comparison methodology}
Both \citeA{leeAutomatedCraterDetection2021} and \citeA{benedixDerivingSurfaceAges2020} compare their results against the \textbf{Robbins} catalog as a source of `ground-truth'. In each paper, the comparison required matching every crater from the CDA catalog with a crater in the \textbf{Robbins} catalog and required the definition of a matching algorithm. The L19 algorithm uses the following criteria to determine a positive crater match to the ground-truth:
\begin{linenomath*}
\begin{eqnarray}
 \nonumber F_\mathrm{D} &= \frac{D_\mathrm{C} - D_\mathrm{G}}{min(D_\mathrm{C},D_\mathrm{G})} &\le 0.25, \\
 \nonumber F_\mathrm{Y} &= \kappa_Y\frac{Y_\mathrm{C} - Y_\mathrm{G}}{min(D_\mathrm{C},D_\mathrm{G})} &\le 0.25, \\
 F_\mathrm{X} &= \kappa_X\frac{X_\mathrm{C} - X_\mathrm{G}}{min(D_\mathrm{C},D_\mathrm{G})} &\le 0.25. \label{eqn:lee}
\end{eqnarray}
\end{linenomath*}
Where $D$,$X$,$Y$ are the diameter (km), longitude (degrees), and latitude (degrees) for the catalog crater($C$) and ground-truth crater ($G$). $\kappa_Y=\frac{2\pi\,R_\mathrm{mars}}{360}$ and $\kappa_X=\kappa_Y \cos{Y_R}$ convert from distance in degrees to distance in kilometers for a spherical planet.

A positive match is identified in L19 only when all three criteria ($F_\mathrm{D}$,$F_\mathrm{Y}$,$F_\mathrm{X}$) are below 0.25. That is, when a candidate crater and ground-truth crater are both within 25\% of the smaller crater's diameter, and the centers of the two craters are within 25\% of the smaller crater's diameter.

Based on \citeA{benedixDerivingSurfaceAges2020}, the B20 method uses the following criteria to determine a positive crater match:
\begin{linenomath*}
\begin{eqnarray}
 \nonumber F_\mathrm{D} &= \frac{D_\mathrm{C} - D_\mathrm{G}}{min(D_\mathrm{C},D_\mathrm{G})} &\le 0.5, \\
 \nonumber F_\mathrm{Y} &= \frac{Y_\mathrm{C} - Y_\mathrm{G}}{|Y_\mathrm{G}|} &\le 0.02, \\
 F_\mathrm{X} &= \frac{X_\mathrm{C} - X_\mathrm{G}}{|X_\mathrm{G}|} &\le 0.02. \label{eqn:benedix}
\end{eqnarray}
\end{linenomath*}
A positive match is identified in B20 if the diameters are within 50\% and the crater centers are within 2\% of their absolute location in longitude or latitude. Note that the B20 method doesn't compare location relative to the diameter as L19 does, using the location in degrees in the denominator instead. B20 also allows matches when the craters do not overlap, while L19 requires an overlap by construction (the center of one crater must be within the circle representing the matching crater).

\subsection{Comparison Metrics}
\label{sec:metrics}
I calculated the performance of the catalogs with respect to the ground-truth catalogs using the standard metric calculations \cite<e.g.,>[]{leeAutomatedCraterDetection2019,Goodfellow2016} that are also used in \citeA{leeAutomatedCraterDetection2021} and \citeA{benedixDerivingSurfaceAges2020}.

Each candidate catalog is constructed from a number $N_C$ total craters, compared to $N_G$ ground-truth craters. Comparing these catalogs identifies a number $N_T$ matching craters (`true positive') and $N_F$ non-matching craters (`false positive'). The performance metrics are all calculated using these crater counts. \textit{Recall} is the fraction of craters in the ground-truth catalog that also exists in the candidate catalog: $R=N_T/N_G$. \textit{Precision} is the fraction of craters in the candidate catalog that exists in the ground-truth catalog $R=N_T/N_C$. Since these metrics measure two distinct properties of the catalog, a blended metric ($F_1$ score) is also typically used, defined as the harmonic mean of \textit{Precision} and \textit{Recall}: $F_1 = 2PR/(P+R)$. 

All three metrics are typically expressed as percentages. For example, \citeA{benedixDerivingSurfaceAges2020} reported their catalog has a \textit{Recall} of 77\% and \textit{Precision} of 75\% (their table 1), \citeA{leeAutomatedCraterDetection2021} reported their catalog has a \textit{Recall} of 80\% and \textit{Precision} of 87\% (their table 1). However, since these catalogs have different bounds on diameter and different spatial coverage, a direct comparison of these numbers is not practical.

For the CDA craters with a matching ground-truth crater (i.e., craters part of the $N_T$ count above), I also calculated the percentage errors between matching pairs of craters based on equations \ref{eqn:benedix} and \ref{eqn:lee} above. The percentage errors are then aggregated and plotted as histograms following figure 3 of \citeA{benedixDerivingSurfaceAges2020}. Since the L19 method calculated errors relative to diameter, these errors cannot be directly compared with the B20 method, but the diameter errors used the same calculation and can be compared.

For each matching crater, I also calculated the Intersection-Over-Union (IOU) metric that measures the overlap (intersection) of two features relative to the total exposed surface area (union) \cite<e.g.,>[]{fairweatherAutomaticMappingSmall2022,benedixDerivingSurfaceAges2020}. The IOU calculation takes the pair of matching craters and calculates the ratio
$$IOU=\frac{\textrm{Total intersecting area of the crater pair}}{\textrm{Total exposed area of the crater pair}}$$
Perfectly overlapping craters have a maximum value of $IOU=1$, while non-overlapping craters have a value of $IOU=0$. Values between 0 and 1 include mismatched sizes and locations but require some overlap. For each of the matched crater pairs, I calculated the IOU metric assuming circular craters since all of the catalogs used in this comparison report crater parameters assuming circular features, even where the underlying craters are degraded or elliptical.

Finally, I calculated the \textit{Recall} and \textit{Precision} as a function of location and size to show the performance of each network across the planet and as a function of crater diameter. In principle, an unbiased network should perform equally well across the planet but will perform worse as the crater size approaches the native image scale and crater features become harder to detect.

Note that the main text of \citeA{benedixDerivingSurfaceAges2020} also describes an `uncorrelated crater' that is calculated in their table 1 as the number of ground-truth craters not found by the CDA (usually called a `false negative'), but it is described as indicating ``where the algorithm found a crater that is not present in the [ground-truth].''. Similarly, their `false positive' craters are described as being found by the ground-truth but not by the CDA (their section 3.3). These definitions are reversed compared to normal usage, and in their \cite{benedixDerivingSurfaceAges2020} supplementary material S3 and their table 1 the calculations follow the standard definitions given above.

\section{Results}\label{section:results}

Four catalogs are used in this section. The \textbf{DeepMars2} catalog from \citeA{leeAutomatedCraterDetection2021}, the \textbf{irH1k} infra-red channel catalog from \citeA{leeAutomatedCraterDetection2021}, the \textbf{Benedix} catalog from \citeA{benedixDerivingSurfaceAges2020}, and my attempted correction to the \textbf{Benedix} catalog in the \textbf{B\_Rescaled} catalog. Table \ref{table1} summarizes the performance of each catalog measured against the \textbf{Robbins} ground-truth catalog. 

The table shows the performance for both the B20 and L19 methodology for craters from 1.5km to 10km in diameter within 65 degrees from the equator (the limits used in \citeA{benedixDerivingSurfaceAges2020}). The values for the \textbf{Benedix} catalog differ by 2-3\% compared to \citeA{benedixDerivingSurfaceAges2020}, but these numbers are sensitive to the numerical algorithm used to match craters. For example, I could increase the \textit{Recall} and \textit{Precision} by 2\% by allowing each ground-truth crater to match multiple candidate craters, but it's not clear that \citeA{benedixDerivingSurfaceAges2020} allowed this to occur.

The performance values for the \textbf{DeepMars2} are consistent with \citeA{leeAutomatedCraterDetection2021}, with a 1\% difference in both \textit{Recall} and \textit{Precision} caused by not removing ${\sim}100$ overlapping craters in the ground-truth catalog for this work.

The B20 and L19 crater matching methods produce very different results. The effect is most noticeable in the \textbf{Benedix} catalog, where performance drops by 35 percentage points using the stricter L19 method (a factor of two in CDA performance), but \textbf{irH1k} also drops by 10\% when changing from the B20 method to the stricter L19 method. In contrast, the \textbf{DeepMars2} is largely unaffected because most non-overlapping craters were removed during post-processing \cite{leeAutomatedCraterDetection2021}.

\begin{table}
 \begin{tabular}{l|l|r|r|r|r|r|r|r}
    \savehline
    Method & Test & R & P & F1 & TP & FP & FN & test count \\
    \savehline
    \multirow{4}{*}{B20} & DeepMars2 & 20 & 95 & 33 & 32,995 & 1,672 & 130,416 & 34,667 \\
     & Benedix & 72 & 70 & 71 & 117,485 & 50,875 & 45,926 & 168,360 \\
     & B\_Rescaled & 72 & 73 & 72 & 117,239 & 43,033 & 46,172 & 160,272 \\
     & irH1k & 76 & 74 & 75 & 124,190 & 44,215 & 39,221 & 168,405 \\
     \savehline
    \multirow{4}{*}{L19} & DeepMars2 & 20 & 92 & 32 & 32,013 & 2,654 & 131,398 & 34,667 \\
     & Benedix & 36 & 35 & 35 & 58,777 & 109,583 & 104,634 & 168,360 \\
     & B\_Rescaled & 47 & 48 & 47 & 76,410 & 83,862 & 87,001 & 160,272 \\
     & irH1k & 65 & 63 & 64 & 105,717 & 62,688 & 57,694 & 168,405 \\
    \savehline
    \end{tabular}
     \caption{ \label{table1} Table of metrics for each catalog with respect to the \citeA{Robbins2012a} catalog for craters within 65 degrees of the equator and between 1.5km and 10km in diameter. Recall, Precision, and F1 are rounded percentages, while all other numeric columns are integer values. A total of 163,411 ground-truth craters were in the comparison. R=Recall, P=Precision, F1=harmonic mean score, TP=true positive count,FP=false positive count, FN=false negative count (maybe listed as \textit{uncorrelated craters} in \citeA{benedixDerivingSurfaceAges2020}), }
\end{table}

\begin{figure}
 \includegraphics[width=0.4\textwidth]{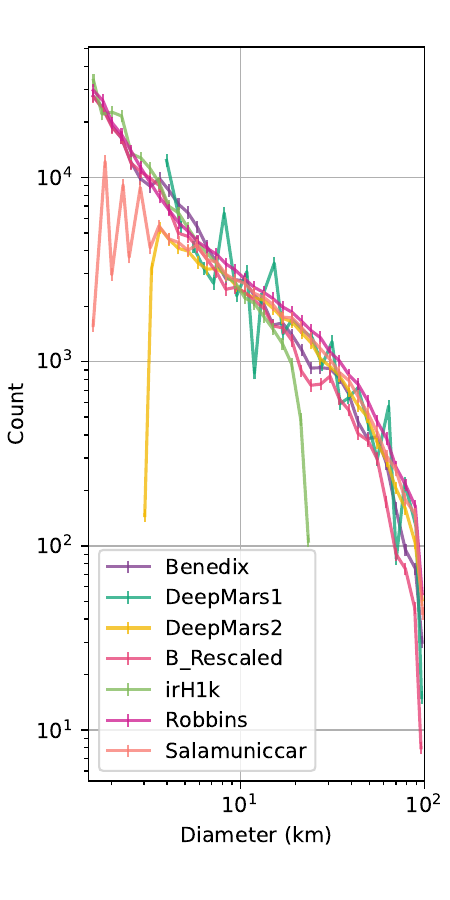}
 \caption{Size Frequency Distribution of all craters in each catalog. 
 \textbf{Benedix} from \citeA{benedixDerivingSurfaceAges2020}, \textbf{DeepMars1} from \citeA{leeAutomatedCraterDetection2019} , \textbf{DeepMars2} from \citeA{leeAutomatedCraterDetection2021}, \textbf{irH1k} from the high resolution IR channel \cite{leeAutomatedCraterDetection2021}, \textbf{Robbins} from \citeA{Robbins2012a}, and \textbf{Salamuniccar} from \citeA{Salamuniccar2012}.\label{figure_csfd}}
\end{figure}

Figure \ref{figure_csfd} shows the size frequency distribution using the same bins as \citeA{benedixDerivingSurfaceAges2020} (20 diameter bins per log-decade). The plot shows that most catalogs have similar distribution shapes, even with different precision and recall metrics. The \citeA{Salamuniccar2012} catalog is truncated above 25km, and it shows discretization artifacts at lower diameters, as does \textbf{DeepMars1} from \citeA{leeAutomatedCraterDetection2019}. The \textbf{irH1k} catalog only processed images up to 1.3 degrees across (a maximum crater size of about 40km).

\begin{figure}
 \includegraphics[width=\textwidth]{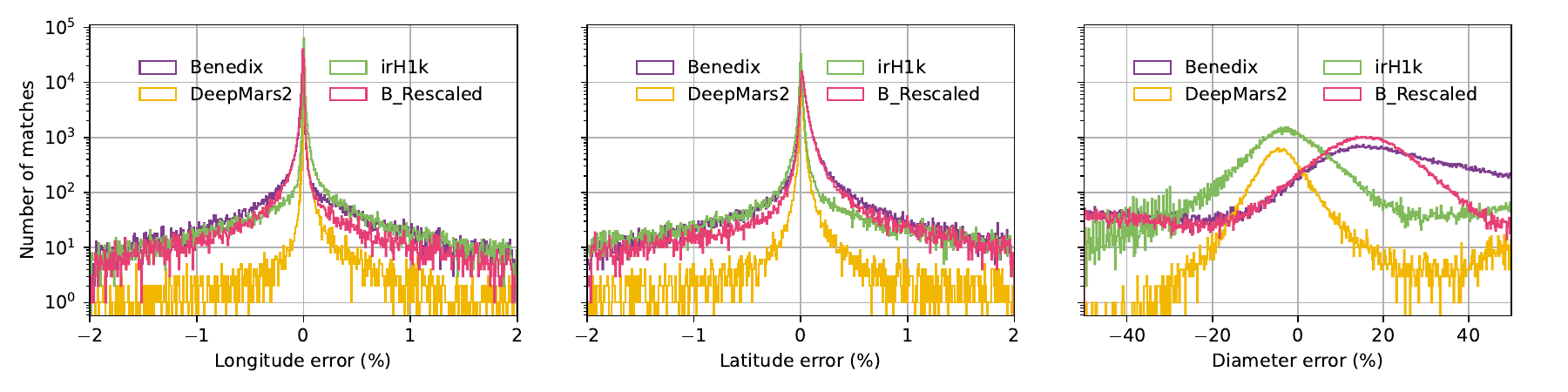}
 \caption{Distribution of errors in longitude (left), latitude (center), and crater Diameter (right) calculated according to the B20 method (equation \ref{eqn:benedix}) using 500 bins in each histogram. Calculations were made relative to the ground-truth catalog such that positive numbers imply a positive bias in the CDA catalog\label{figure_benedix}.}
\end{figure}

Figure \ref{figure_benedix} shows the distribution of errors for craters matched between a candidate catalog and the \textbf{Robbins} catalog using the B20 method (equation \ref{eqn:benedix}). The shapes of these figures agree reasonably well with \citeA{benedixDerivingSurfaceAges2020}. The location error is relatively unbiased with a prominent central peak, and the diameter errors shown here match the reported positive bias\cite{benedixDerivingSurfaceAges2020}. The \textbf{DeepMars2} catalog has a lower total count but similar biases to the \textbf{irH1K} catalog. The diameter biases have different signs between the \citeA{benedixDerivingSurfaceAges2020} and \citeA{leeAutomatedCraterDetection2021} catalogs, possibly due to a difference in network training.

\begin{figure}
 \includegraphics[width=\textwidth]{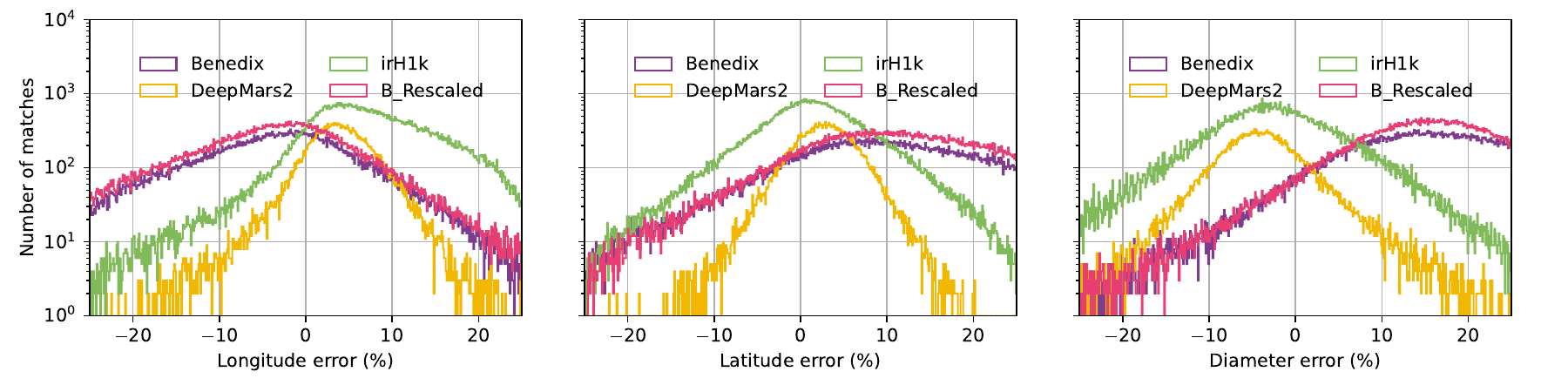}
 \caption{Distribution of errors in longitude (left), latitude (center), and crater Diameter (right) calculated according to the L19 method (equation \ref{eqn:lee}). The horizontal extent of the longitude and latitude plot appear larger here but are expressed relative to the crater diameter rather than the absolute value of longitude or latitude used in figure \ref{figure_benedix}. Calculations were made relative to the ground-truth catalog such that positive numbers imply a positive bias in the CDA catalog\label{figure_lee}}
\end{figure}

Figure \ref{figure_lee} shows the calculation of errors using the L19 method (equation \ref{eqn:lee}). Note that all errors are now calculated relative to the diameter of the smaller crater in a matched pair since the L19 method included that diameter as the denominator in each calculation. The shape of the diameter plot is essentially the core of the same subplot in figure \ref{figure_benedix}. In contrast, the longitude and latitude figures are functionally different, showing the position error \textit{relative} to the crater diameter rather than in \textit{absolute} location as in figure \ref{figure_benedix}. 

\begin{figure}
 \includegraphics[width=0.5\textwidth]{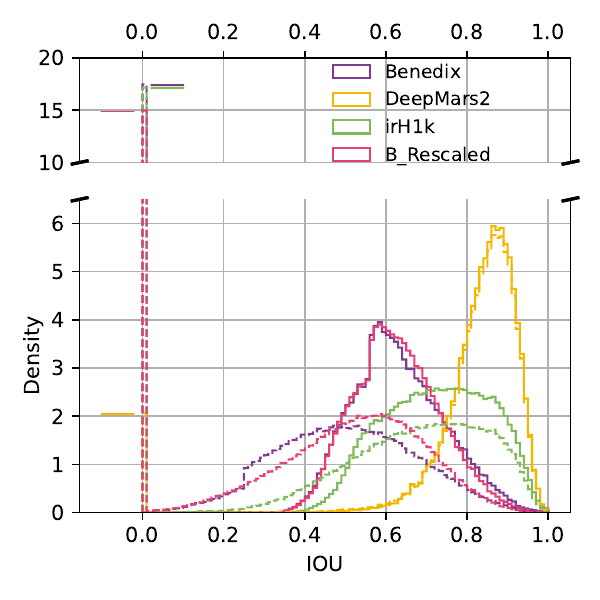}
 \caption{Intersection-over-Union values for the four crater catalogs. Solid lines show the IOUs of crater pairs matched using the L19 method. Dashed lines show the same catalogs matched using the B20 method, and horizontal lines at IOU=0 highlight the population of non-overlapping craters ($IOU=0$) found using the B20 method. The histograms used 100 bins.\label{figure_iou}}
\end{figure}

Figure \ref{figure_iou} shows the distribution of Intersection-over-Union (IOU) values for matching craters using both methods (L19 in solid lines, B20 in dashed lines) and for each of catalogs relative to \textbf{Robbins}. We did not calculate IOU metrics in training the L19 or L21 CDA because the algorithm matched crater rims with narrow widths. \citeA{benedixDerivingSurfaceAges2020} reported that validation and training were based on IOU comparisons with the ground truth and required a high threshold for a positive detection. The plot shows $IOU>0.3$ for all matches in the L19 method (solid lines) compared to the theoretical minimum of $IOU\approx0.25$ given the constraints in equation \ref{eqn:lee}. Since the B20 method counts some non-overlapping craters as matches, there is a population of matching craters with no intersection and a value $IOU=0$, marked with a horizontal line for each catalog at the left edge of the plot.

\begin{figure}
 \includegraphics[width=\textwidth]{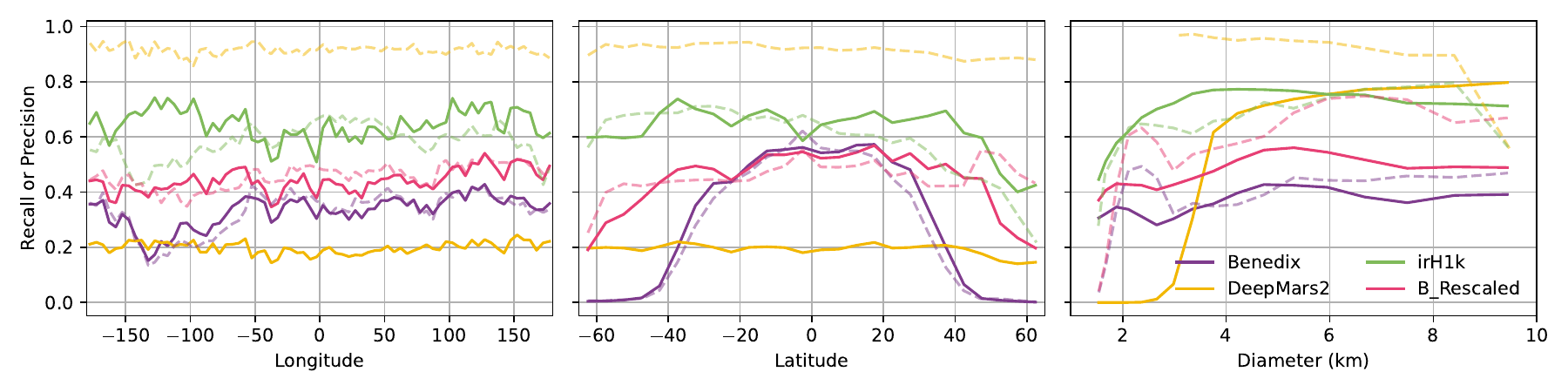}
 \caption{\textit{Recall} (solid) and \textit{Precision} (dashed) plotted as a function of longitude, latitude, and diameter for catalogs matched using the L19 method. Longitude and latitude are aggregated into 5-degree bins, and diameter is aggregated into 20 bins per log-decade\label{figure_recall}.}
\end{figure}

Figure \ref{figure_recall} shows the \textit{Recall} and \textit{Precision} metrics as a function of longitude, latitude, and diameter based on matched-craters using the L19 method. The \textbf{DeepMars2} catalog has the largest difference between \textit{Recall} and \textit{Precision} because it removes all craters below 3.5km in diameter and thus misses 100,000 small craters, lowering the \textit{Recall} across all longitudes and latitudes. All four catalogs have consistent performance in longitude, fluctuating about 10\% overall. In latitude, the \textbf{Benedix} catalog has a significant performance drop with increasing latitude in both hemispheres, while the \textbf{irH1k} has a small drop in performance with latitude with worse performance in the northern hemisphere. All four catalogs lose performance as the crater diameter approaches the pixel scale of the source imagery, as expected. For \textbf{DeepMars2}, this drop-off occurs at 3.5km. For the other (IR) catalogs, the drop-off occurs near 2km.

\section{Discussion}\label{section:discussion}
 To generate the comparisons, I needed to make several assumptions to reproduce the B20 method and crater-matching catalogs. Neither \textbf{Benedix} nor \textbf{DeepMars2} include the corresponding craters from \textbf{Robbins}. Instead, I built the crater-matching algorithms following the description in \citeA{benedixDerivingSurfaceAges2020} (B20, equation \ref{eqn:benedix} above) and \citeA{leeAutomatedCraterDetection2019} (L19, equation \ref{eqn:lee} above). 
 
 Using these two comparison methods, I identified craters that matched with a unique ground-truth crater and used these matches to generate the statistics presented here. The results in table \ref{table1} deviate by a few percent compared to \citeA{benedixDerivingSurfaceAges2020}, but these numbers can be adjusted by a percent or two in either direction by changing the interpretation of parameters in the matching algorithm. Still, the metrics are comparable within this work because they are all calculated with the same methodology and code. The SFD (figure \ref{figure_csfd}) and the error distribution (figure \ref{figure_benedix}) match well with the corresponding figures from \citeA{benedixDerivingSurfaceAges2020}, providing some confidence that the implementation of the B20 method matches the original. Similarly, the metrics and biases reported in \citeA{leeAutomatedCraterDetection2021} agree with the values calculated here using the L19 method.

The B20 algorithm produced higher apparent performance because it included craters that don't overlap with their targets, accounting for 50\% of \textbf{Benedix} matches, 3\% of \textbf{DeepMars2}, and 16\% of \textbf{irH1k}. The L19 algorithm required overlapping craters for a valid match and thus produced higher IOU distributions in figure \ref{figure_iou} but lower overall performance. Craters that matched according to L19 also matched according to B20 except for the few craters along the equator or prime meridian where 2\% of absolute longitude is smaller than 25\% of the crater diameter.
 
 The performance of each catalog is reasonably consistent with longitude and has the expected drop as the crater diameter approaches the image pixel scale. There is a drop in performance with latitude in the catalogs, but a much larger drop in performance with latitude in \textbf{Benedix} suggests a more fundamental error caused by neglecting to project images onto orthographic projections before processing.

 The source of the error is not obvious from the available catalogs without access to the processing algorithms. However, of the 58,578 \textbf{Benedix} catalog craters matched by B20 but not by L19, only 13,383 would satisfy the diameter constraints of the L19 method. If the crater diameters in \textbf{Benedix} are affected by a projection error, or equivalently using the wrong pixel-to-kilometer scaling factor, they can be corrected in the existing dataset because the catalog includes pixel scale parameters and location information. Scaling the diameters by cosine latitude improves the catalog performance using the L19 method by allowing an additional 23,309 craters to match the ground-truth catalog.
 
\begin{figure}
 \includegraphics[width=0.4\textwidth]{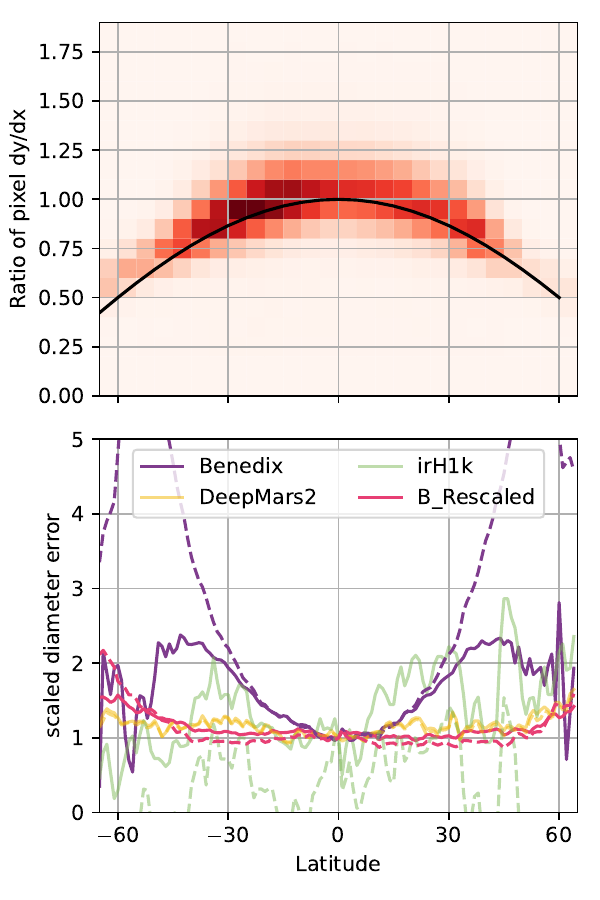}
 \caption{(top) Heat map of crater population as a function of latitude and the ratio of height-to-width (dy/dx) of the pixel-level rectangle in the \textbf{Benedix} catalog. Darker shades represent a higher density of craters, and the black line shows the cosine of latitude. (bottom) Scaled diameter bias as a function of latitude, calculated as the median $F_D$ from equation \ref{eqn:lee} (solid) or equation \ref{eqn:benedix} (dashed) relative to its value at the equator. The drop off in apparent bias at high latitude is because the high biases exceed the threshold in each method, and the remaining craters have lower bias\label{figure_cosine}.}
 \end{figure}

 This hypothesis is supported by the raw pixel size information provided in the \textbf{Benedix} catalog. Figure \ref{figure_cosine}(top) shows the density map of craters as a function of the latitude and the ratio of the height (`dy') to width (`dx') of the rectangle encompassing the crater in pixel values. This ratio should be near 1 for orthographically projected images for circular craters at all latitudes, but has a cosine latitude dependence for Plate Car\`ee (unprojected, or latitude-longitude) images. In unprojected images, circular craters appear stretched in longitude at higher latitudes, and the ratio dy/dx decreases for circular features, as shown in the figure. This issue suggests the image swatches were presented to the CDA as unprojected images where circular physical features appear elliptical in the images, and the CDA was trained to find horizontally stretched ellipses rather than circles.
 
 The pixel level information can also be used to determine how the conversion to geophysical units (in this case, kilometers) was done by \citeA{benedixDerivingSurfaceAges2020}. Three image scales were used by \citeA{benedixDerivingSurfaceAges2020}, and these scales accurately map all of the report sizes in the `x' dimension from the pixel size to the reported physical size of the crater in kilometers. However, in these images, `x' corresponds to longitude and should have a different scaling with latitude, whereas the `y' axis corresponds to the pixel scale in latitude and does not vary with location. Thus, the crater's pixel size in the `y' axis should have been used instead of the size in `x'. As a result, the \textbf{Benedix} catalog has an increasing positive bias in crater diameter as a function of latitude (figure \ref{figure_cosine}(bottom)).

 My proposed fix for this error is to apply the image pixel scaling to the `dy' pixel values instead of `dx', which produces the \textbf{B\_Rescaled} catalog shown in each table and figure. This new catalog has improved global metrics, reduced errors, and increased IOU values relative to \textbf{Benedix}. The \textit{Recall} and \textit{Precision} metrics also remain higher with latitude and increase in all dimensions compared to the original catalog. However, this improvement does not definitively prove that the \textbf{Benedix} catalog has a projection issue, and other issues might cause the observed bias.
 
\section{Conclusions}\label{section:conclusions}
I have examined crater catalogs generated with neural networks \cite{leeAutomatedCraterDetection2021,benedixDerivingSurfaceAges2020} whose results suggested similar performance but were calculated using different methodologies. Using the comparison method from \citeA{leeAutomatedCraterDetection2019} I found that the \citeA{benedixDerivingSurfaceAges2020} catalog has significantly lower \textit{Precision} compared to \citeA{leeAutomatedCraterDetection2021} generating 30 times more false crater detections, but did contain more craters in total and with a higher \textit{Recall}. When compared to the infrared channel of \citeA{leeAutomatedCraterDetection2021} (the \textbf{irH1k} catalog) that found craters down to 1km, the \citeA{benedixDerivingSurfaceAges2020} catalog has lower performance in all metrics.

I also identified a possible scaling error in the \citeA{benedixDerivingSurfaceAges2020} catalog related either to image projection or using the incorrect pixel-to-kilometer scaling constant. I proposed a correction that rescaled to crater sizes by the correct scaling coefficient and generated the \textbf{B\_Rescaled} catalog. This rescaled catalog improved in all metrics over the original catalog.

Since the validation of crater catalogs becomes increasingly difficult with their increasing size, the ground-truth comparison methods must be clearly defined and repeatable without ambiguity. In \citeA{leeAutomatedCraterDetection2019}, I attempted to provide enough information to reconstruct the catalogs and provided detailed metric calculations and source code \cite{leeCraterCatalogsSoftware2018}. This catalog also included the matching craters from \citeA{Robbins2012a}. \citeA{leeAutomatedCraterDetection2021} also provided software and catalogs necessary to repeat the analysis \citeA{leeAutomatedCraterDetection2021}. Still, we did not identify the matching ground-truth craters or the `true positives' from CDA catalogs. Similarly, \citeA{benedixDerivingSurfaceAges2020} provided an accessible catalog of craters but with no software or identifications of ground-truth craters, and a more narrative description of the comparison methods.

Since these CDAs are now used to study craters smaller than existing catalog limits, there are no ground-truth catalogs. It has become necessary to use independent neural networks or other tools to provide supporting validation of new catalogs. In \citeA{leeAutomatedCraterDetection2021}, we used multiple networks trained on different input datasets to increase the overall performance. In \citeA{Lee2018a}, I used multiple networks on the same high-resolution DTM dataset to achieve the same result. In \citeA{lagainMarsCraterDatabase2021}, new web-based tools were developed to validate an existing catalog and found several thousand new craters and possible mis-identifications. These additional validation tools from secondary networks or manual validation efforts should be considered necessary in future studies to provide confidence in new crater catalogs.

Finally, since algorithm errors can arise, the source code and detailed methodology for the entire CDA should be an essential part of future research contributions. The source code and processed catalogs used in the analysis here are provided at \citeA{leeReplicationDataComparison2023}.

\section*{Open Research Section}
The two published crater catalogs compared here can be found at \citeA{benedixDerivingSurfaceAges2020} and \citeA{leeAutomatedCraterDetection2021} and links therein. Software and catalog comparison data for this work is available at \url{https://doi.org/10.5683/SP3/AUJMHR}.

\acknowledgments
The author thanks \citeA{benedixDerivingSurfaceAges2020} and \citeA{Robbins2012a} for releasing their crater catalogs in easily accessible formats. The Natural Sciences and Engineering Research Council of Canada funded part of this project. The author thanks S. J. Robbins and an anonymous reviewer for valuable comments that help clarify the article.

\bibliography{benedix_comparison}

\begin{thebibliography}{}

\bibitem [\protect \citeauthoryear {%
Benedix%
\ \protect \BOthers {.}}{%
Benedix%
\ \protect \BOthers {.}}{%
{\protect \APACyear {2020}}%
}]{%
benedixDerivingSurfaceAges2020}
\APACinsertmetastar {%
benedixDerivingSurfaceAges2020}%
\begin{APACrefauthors}%
Benedix, G\BPBI K.%
, Lagain, A.%
, Chai, K.%
, Meka, S.%
, Anderson, S.%
, Norman, C.%
\BDBL {}Tan, T.%
\end{APACrefauthors}%
\unskip\
\newblock
\APACrefYearMonthDay{2020}{{\APACmonth{03}}}{}.
\newblock
{\BBOQ}\APACrefatitle {Deriving {{Surface Ages}} on {{Mars Using Automated
  Crater Counting}}} {Deriving {{Surface Ages}} on {{Mars Using Automated
  Crater Counting}}}.{\BBCQ}
\newblock
\APACjournalVolNumPages{Earth and Space Science}{7}{3}{}.
\newblock
\begin{APACrefDOI} \doi{10.1029/2019EA001005} \end{APACrefDOI}
\PrintBackRefs{\CurrentBib}

\bibitem [\protect \citeauthoryear {%
Diakogiannis%
, Waldner%
, Caccetta%
\BCBL {}\ \BBA {} Wu%
}{%
Diakogiannis%
\ \protect \BOthers {.}}{%
{\protect \APACyear {2019}}%
}]{%
Diakogiannis2019}
\APACinsertmetastar {%
Diakogiannis2019}%
\begin{APACrefauthors}%
Diakogiannis, F\BPBI I.%
, Waldner, F.%
, Caccetta, P.%
\BCBL {}\ \BBA {} Wu, C.%
\end{APACrefauthors}%
\unskip\
\newblock
\APACrefYearMonthDay{2019}{}{}.
\newblock
{\BBOQ}\APACrefatitle {{{ResUNet-a}}: A Deep Learning Framework for Semantic
  Segmentation of Remotely Sensed Data} {{{ResUNet-a}}: A deep learning
  framework for semantic segmentation of remotely sensed data}.{\BBCQ}
\newblock
\APACjournalVolNumPages{Arxiv}{}{}{}.
\PrintBackRefs{\CurrentBib}

\bibitem [\protect \citeauthoryear {%
Fairweather%
\ \protect \BOthers {.}}{%
Fairweather%
\ \protect \BOthers {.}}{%
{\protect \APACyear {2022}}%
}]{%
fairweatherAutomaticMappingSmall2022}
\APACinsertmetastar {%
fairweatherAutomaticMappingSmall2022}%
\begin{APACrefauthors}%
Fairweather, J\BPBI H.%
, Lagain, A.%
, Servis, K.%
, Benedix, G\BPBI K.%
, Kumar, S\BPBI S.%
\BCBL {}\ \BBA {} Bland, P\BPBI A.%
\end{APACrefauthors}%
\unskip\
\newblock
\APACrefYearMonthDay{2022}{{\APACmonth{07}}}{}.
\newblock
{\BBOQ}\APACrefatitle {Automatic {{Mapping}} of {{Small Lunar Impact Craters
  Using LRO}}-{{NAC Images}}} {Automatic {{Mapping}} of {{Small Lunar Impact
  Craters Using LRO}}-{{NAC Images}}}.{\BBCQ}
\newblock
\APACjournalVolNumPages{Earth and Space Science}{9}{7}{}.
\newblock
\begin{APACrefDOI} \doi{10.1029/2021EA002177} \end{APACrefDOI}
\PrintBackRefs{\CurrentBib}

\bibitem [\protect \citeauthoryear {%
Giannakis%
, Bhardwaj%
, Sam%
\BCBL {}\ \BBA {} Leontidis%
}{%
Giannakis%
\ \protect \BOthers {.}}{%
{\protect \APACyear {2023}}%
}]{%
giannakisDeepLearningUniversal2023}
\APACinsertmetastar {%
giannakisDeepLearningUniversal2023}%
\begin{APACrefauthors}%
Giannakis, I.%
, Bhardwaj, A.%
, Sam, L.%
\BCBL {}\ \BBA {} Leontidis, G.%
\end{APACrefauthors}%
\unskip\
\newblock
\APACrefYearMonthDay{2023}{{\APACmonth{04}}}{}.
\newblock
\APACrefbtitle {Deep Learning Universal Crater Detection Using {{Segment
  Anything Model}} ({{SAM}})} {Deep learning universal crater detection using
  {{Segment Anything Model}} ({{SAM}})}\ (\BNUM\ arXiv:2304.07764).
\newblock
\APACaddressPublisher{}{{arXiv}}.
\newblock
\begin{APACrefDOI} \doi{10.48550/arXiv.2304.07764} \end{APACrefDOI}
\PrintBackRefs{\CurrentBib}

\bibitem [\protect \citeauthoryear {%
Goodfellow%
, Bengio%
\BCBL {}\ \BBA {} Courville%
}{%
Goodfellow%
\ \protect \BOthers {.}}{%
{\protect \APACyear {2016}}%
}]{%
Goodfellow2016}
\APACinsertmetastar {%
Goodfellow2016}%
\begin{APACrefauthors}%
Goodfellow, I.%
, Bengio, Y.%
\BCBL {}\ \BBA {} Courville, A.%
\end{APACrefauthors}%
\unskip\
\newblock
\APACrefYear{2016}.
\newblock
\APACrefbtitle {Deep {{Learning}}} {Deep {{Learning}}}.
\newblock
\APACaddressPublisher{}{{MIT Press}}.
\PrintBackRefs{\CurrentBib}

\bibitem [\protect \citeauthoryear {%
Lagain%
, Benedix%
\BCBL {}\ \protect \BOthers {.}}{%
Lagain%
, Benedix%
\BCBL {}\ \protect \BOthers {.}}{%
{\protect \APACyear {2021}}%
}]{%
lagainTharsisMantleSource2021}
\APACinsertmetastar {%
lagainTharsisMantleSource2021}%
\begin{APACrefauthors}%
Lagain, A.%
, Benedix, G\BPBI K.%
, Servis, K.%
, Baratoux, D.%
, Doucet, L\BPBI S.%
, Raj{\v s}ic, A.%
\BDBL {}Miljkovi{\'c}, K.%
\end{APACrefauthors}%
\unskip\
\newblock
\APACrefYearMonthDay{2021}{{\APACmonth{11}}}{}.
\newblock
{\BBOQ}\APACrefatitle {The {{Tharsis}} Mantle Source of Depleted Shergottites
  Revealed by 90 Million Impact Craters} {The {{Tharsis}} mantle source of
  depleted shergottites revealed by 90 million impact craters}.{\BBCQ}
\newblock
\APACjournalVolNumPages{Nature Communications}{12}{1}{6352}.
\newblock
\begin{APACrefDOI} \doi{10.1038/s41467-021-26648-3} \end{APACrefDOI}
\PrintBackRefs{\CurrentBib}

\bibitem [\protect \citeauthoryear {%
Lagain%
, Bouley%
\BCBL {}\ \protect \BOthers {.}}{%
Lagain%
, Bouley%
\BCBL {}\ \protect \BOthers {.}}{%
{\protect \APACyear {2021}}%
}]{%
lagainMarsCraterDatabase2021}
\APACinsertmetastar {%
lagainMarsCraterDatabase2021}%
\begin{APACrefauthors}%
Lagain, A.%
, Bouley, S.%
, Baratoux, D.%
, Marmo, C.%
, Costard, F.%
, Delaa, O.%
\BDBL {}Gamblin, O.%
\end{APACrefauthors}%
\unskip\
\newblock
\APACrefYearMonthDay{2021}{{\APACmonth{08}}}{}.
\newblock
{\BBOQ}\APACrefatitle {Mars {{Crater Database}}: {{A}} Participative Project
  for the Classification of the Morphological Characteristics of Large
  {{Martian}} Craters} {Mars {{Crater Database}}: {{A}} participative project
  for the classification of the morphological characteristics of large
  {{Martian}} craters}.{\BBCQ}
\newblock
\BIn{} W\BPBI U.~Reimold\ \BBA {} C.~Koeberl\ (\BEDS), \APACrefbtitle {Large
  {{Meteorite Impacts}} and {{Planetary Evolution VI}}} {Large {{Meteorite
  Impacts}} and {{Planetary Evolution VI}}}\ (\BVOL~550, \BPG~0).
\newblock
\APACaddressPublisher{}{{Geological Society of America}}.
\newblock
\begin{APACrefDOI} \doi{10.1130/2021.2550(29)} \end{APACrefDOI}
\PrintBackRefs{\CurrentBib}

\bibitem [\protect \citeauthoryear {%
Latorre%
, Spiller%
, Sasidharan%
, Basheer%
\BCBL {}\ \BBA {} Curti%
}{%
Latorre%
\ \protect \BOthers {.}}{%
{\protect \APACyear {2023}}%
}]{%
latorreTransferLearningRealtime2023}
\APACinsertmetastar {%
latorreTransferLearningRealtime2023}%
\begin{APACrefauthors}%
Latorre, F.%
, Spiller, D.%
, Sasidharan, S\BPBI T.%
, Basheer, S.%
\BCBL {}\ \BBA {} Curti, F.%
\end{APACrefauthors}%
\unskip\
\newblock
\APACrefYearMonthDay{2023}{{\APACmonth{04}}}{}.
\newblock
{\BBOQ}\APACrefatitle {Transfer Learning for Real-Time Crater Detection on
  Asteroids Using a {{Fully Convolutional Neural Network}}} {Transfer learning
  for real-time crater detection on asteroids using a {{Fully Convolutional
  Neural Network}}}.{\BBCQ}
\newblock
\APACjournalVolNumPages{Icarus}{394}{}{115434}.
\newblock
\begin{APACrefDOI} \doi{10.1016/j.icarus.2023.115434} \end{APACrefDOI}
\PrintBackRefs{\CurrentBib}

\bibitem [\protect \citeauthoryear {%
Lee%
}{%
Lee%
}{%
{\protect \APACyear {2018}}%
{\protect \APACexlab {{\protect \BCnt {1}}}}}]{%
leeCraterCatalogsSoftware2018}
\APACinsertmetastar {%
leeCraterCatalogsSoftware2018}%
\begin{APACrefauthors}%
Lee, C.%
\end{APACrefauthors}%
\unskip\
\newblock
\APACrefYearMonthDay{2018{\protect \BCnt {1}}}{}{}.
\newblock
\APACrefbtitle {Crater {{Catalogs}} and Software for "{{Automated}} Crater
  Detection on {{Mars}} Using Deep Learning".} {Crater {{Catalogs}} and
  software for "{{Automated}} crater detection on {{Mars}} using deep
  learning".}
\PrintBackRefs{\CurrentBib}

\bibitem [\protect \citeauthoryear {%
Lee%
}{%
Lee%
}{%
{\protect \APACyear {2018}}%
{\protect \APACexlab {{\protect \BCnt {2}}}}}]{%
Lee2018a}
\APACinsertmetastar {%
Lee2018a}%
\begin{APACrefauthors}%
Lee, C.%
\end{APACrefauthors}%
\unskip\
\newblock
\APACrefYearMonthDay{2018{\protect \BCnt {2}}}{}{}.
\newblock
{\BBOQ}\APACrefatitle {Martian {{Crater Identification Using Deep Learning}}}
  {Martian {{Crater Identification Using Deep Learning}}}.{\BBCQ}
\newblock
\BIn{} \APACrefbtitle {American {{Geophysical Union Fall Meeting}}} {American
  {{Geophysical Union Fall Meeting}}}\ (\BPG~P41D-3768).
\PrintBackRefs{\CurrentBib}

\bibitem [\protect \citeauthoryear {%
Lee%
}{%
Lee%
}{%
{\protect \APACyear {2019}}%
}]{%
leeAutomatedCraterDetection2019}
\APACinsertmetastar {%
leeAutomatedCraterDetection2019}%
\begin{APACrefauthors}%
Lee, C.%
\end{APACrefauthors}%
\unskip\
\newblock
\APACrefYearMonthDay{2019}{}{}.
\newblock
{\BBOQ}\APACrefatitle {Automated Crater Detection on {{Mars}} Using {{Deep
  Learning}}} {Automated crater detection on {{Mars}} using {{Deep
  Learning}}}.{\BBCQ}
\newblock
\APACjournalVolNumPages{Planetary and Space Science}{}{}{}.
\newblock
\begin{APACrefDOI} \doi{10.1016/j.pss.2019.03.008} \end{APACrefDOI}
\PrintBackRefs{\CurrentBib}

\bibitem [\protect \citeauthoryear {%
Lee%
}{%
Lee%
}{%
{\protect \APACyear {2023}}%
{\protect \APACexlab {{\protect \BCnt {1}}}}}]{%
leeReplicationDataComparison}
\APACinsertmetastar {%
leeReplicationDataComparison}%
\begin{APACrefauthors}%
Lee, C.%
\end{APACrefauthors}%
\unskip\
\newblock
\APACrefYearMonthDay{2023{\protect \BCnt {1}}}{}{}.
\newblock
\APACrefbtitle {Replication {{Data}} for: {{Comparison}} of Automated Crater
  Catalogs for {{Mars}} from {{Benedix}} et al. (2020) and {{Lee}} and
  {{Hogan}} (2021).} {Replication {{Data}} for: {{Comparison}} of automated
  crater catalogs for {{Mars}} from {{Benedix}} et al. (2020) and {{Lee}} and
  {{Hogan}} (2021).}
\newblock
\APACaddressPublisher{{borealisdata.ca}}{{Dataverse}}.
\newblock
\begin{APACrefDOI} \doi{10.5683/SP3/AUJMHR} \end{APACrefDOI}
\PrintBackRefs{\CurrentBib}

\bibitem [\protect \citeauthoryear {%
Lee%
}{%
Lee%
}{%
{\protect \APACyear {2023}}%
{\protect \APACexlab {{\protect \BCnt {2}}}}}]{%
leeReplicationDataComparison2023}
\APACinsertmetastar {%
leeReplicationDataComparison2023}%
\begin{APACrefauthors}%
Lee, C.%
\end{APACrefauthors}%
\unskip\
\newblock
\APACrefYearMonthDay{2023{\protect \BCnt {2}}}{{\APACmonth{07}}}{}.
\newblock
\APACrefbtitle {Replication {{Data}} for: {{Comparison}} of Automated Crater
  Catalogs for {{Mars}} from {{Benedix}} et al. (2020) and {{Lee}} and
  {{Hogan}} (2021).} {Replication {{Data}} for: {{Comparison}} of automated
  crater catalogs for {{Mars}} from {{Benedix}} et al. (2020) and {{Lee}} and
  {{Hogan}} (2021).}
\newblock
\APACaddressPublisher{}{{Borealis}}.
\newblock
\begin{APACrefDOI} \doi{10.5683/SP3/AUJMHR} \end{APACrefDOI}
\PrintBackRefs{\CurrentBib}

\bibitem [\protect \citeauthoryear {%
Lee%
\ \BBA {} Hogan%
}{%
Lee%
\ \BBA {} Hogan%
}{%
{\protect \APACyear {2020}}%
}]{%
leeCraterCatalogsSoftware2020}
\APACinsertmetastar {%
leeCraterCatalogsSoftware2020}%
\begin{APACrefauthors}%
Lee, C.%
\BCBT {}\ \BBA {} Hogan, J.%
\end{APACrefauthors}%
\unskip\
\newblock
\APACrefYearMonthDay{2020}{}{}.
\newblock
\APACrefbtitle {Crater {{Catalogs}} and Software for "{{Automated}} Crater
  Detection with Human Level Performance".} {Crater {{Catalogs}} and software
  for "{{Automated}} crater detection with human level performance".}
\newblock
\begin{APACrefDOI} \doi{10.5683/SP2/CFUNII} \end{APACrefDOI}
\PrintBackRefs{\CurrentBib}

\bibitem [\protect \citeauthoryear {%
Lee%
\ \BBA {} Hogan%
}{%
Lee%
\ \BBA {} Hogan%
}{%
{\protect \APACyear {2021}}%
}]{%
leeAutomatedCraterDetection2021}
\APACinsertmetastar {%
leeAutomatedCraterDetection2021}%
\begin{APACrefauthors}%
Lee, C.%
\BCBT {}\ \BBA {} Hogan, J.%
\end{APACrefauthors}%
\unskip\
\newblock
\APACrefYearMonthDay{2021}{{\APACmonth{02}}}{}.
\newblock
{\BBOQ}\APACrefatitle {Automated Crater Detection with Human Level Performance}
  {Automated crater detection with human level performance}.{\BBCQ}
\newblock
\APACjournalVolNumPages{Computers and Geosciences}{147}{}{104645}.
\newblock
\begin{APACrefDOI} \doi{10.1016/j.cageo.2020.104645} \end{APACrefDOI}
\PrintBackRefs{\CurrentBib}

\bibitem [\protect \citeauthoryear {%
Malin%
\ \protect \BOthers {.}}{%
Malin%
\ \protect \BOthers {.}}{%
{\protect \APACyear {2007}}%
}]{%
malinContextCameraInvestigation2007}
\APACinsertmetastar {%
malinContextCameraInvestigation2007}%
\begin{APACrefauthors}%
Malin, M\BPBI C.%
, Bell~III, J\BPBI F.%
, Cantor, B\BPBI A.%
, Caplinger, M\BPBI A.%
, Calvin, W\BPBI M.%
, Clancy, R\BPBI T.%
\BDBL {}Wolff, M\BPBI J.%
\end{APACrefauthors}%
\unskip\
\newblock
\APACrefYearMonthDay{2007}{}{}.
\newblock
{\BBOQ}\APACrefatitle {Context {{Camera Investigation}} on Board the {{Mars
  Reconnaissance Orbiter}}} {Context {{Camera Investigation}} on board the
  {{Mars Reconnaissance Orbiter}}}.{\BBCQ}
\newblock
\APACjournalVolNumPages{Journal of Geophysical Research: Planets}{112}{E5}{}.
\newblock
\begin{APACrefDOI} \doi{10.1029/2006JE002808} \end{APACrefDOI}
\PrintBackRefs{\CurrentBib}

\bibitem [\protect \citeauthoryear {%
Redmon%
, Divvala%
, Girshick%
\BCBL {}\ \BBA {} Farhadi%
}{%
Redmon%
\ \protect \BOthers {.}}{%
{\protect \APACyear {2016}}%
}]{%
redmonYouOnlyLook2016}
\APACinsertmetastar {%
redmonYouOnlyLook2016}%
\begin{APACrefauthors}%
Redmon, J.%
, Divvala, S.%
, Girshick, R.%
\BCBL {}\ \BBA {} Farhadi, A.%
\end{APACrefauthors}%
\unskip\
\newblock
\APACrefYearMonthDay{2016}{{\APACmonth{06}}}{}.
\newblock
{\BBOQ}\APACrefatitle {You {{Only Look Once}}: {{Unified}}, {{Real-Time Object
  Detection}}} {You {{Only Look Once}}: {{Unified}}, {{Real-Time Object
  Detection}}}.{\BBCQ}
\newblock
\BIn{} \APACrefbtitle {2016 {{IEEE Conference}} on {{Computer Vision}} and
  {{Pattern Recognition}} ({{CVPR}})} {2016 {{IEEE Conference}} on {{Computer
  Vision}} and {{Pattern Recognition}} ({{CVPR}})}\ (\BPGS\ 779--788).
\newblock
\begin{APACrefDOI} \doi{10.1109/CVPR.2016.91} \end{APACrefDOI}
\PrintBackRefs{\CurrentBib}

\bibitem [\protect \citeauthoryear {%
Redmon%
\ \BBA {} Farhadi%
}{%
Redmon%
\ \BBA {} Farhadi%
}{%
{\protect \APACyear {2018}}%
}]{%
redmonYOLOv3IncrementalImprovement2018}
\APACinsertmetastar {%
redmonYOLOv3IncrementalImprovement2018}%
\begin{APACrefauthors}%
Redmon, J.%
\BCBT {}\ \BBA {} Farhadi, A.%
\end{APACrefauthors}%
\unskip\
\newblock
\APACrefYearMonthDay{2018}{{\APACmonth{04}}}{}.
\newblock
\APACrefbtitle {{{YOLOv3}}: {{An Incremental Improvement}}} {{{YOLOv3}}: {{An
  Incremental Improvement}}}\ (\BNUM\ arXiv:1804.02767).
\newblock
\APACaddressPublisher{}{{arXiv}}.
\newblock
\begin{APACrefDOI} \doi{10.48550/arXiv.1804.02767} \end{APACrefDOI}
\PrintBackRefs{\CurrentBib}

\bibitem [\protect \citeauthoryear {%
Robbins%
}{%
Robbins%
}{%
{\protect \APACyear {2017}}%
}]{%
Robbins2017}
\APACinsertmetastar {%
Robbins2017}%
\begin{APACrefauthors}%
Robbins, S\BPBI J.%
\end{APACrefauthors}%
\unskip\
\newblock
\APACrefYearMonthDay{2017}{}{}.
\newblock
{\BBOQ}\APACrefatitle {Progress on Old and New: {{D}} {$>$} 1km Crater Catalogs
  for {{Mars}} and {{Moon}}.} {Progress on old and new: {{D}} {$>$} 1km crater
  catalogs for {{Mars}} and {{Moon}}.}{\BBCQ}
\newblock
\BIn{} \APACrefbtitle {Planetary {{Crater Consortium}} 8} {Planetary {{Crater
  Consortium}} 8}\ (\BPGS\ 1--2).
\newblock
\begin{APACrefDOI} \doi{10.1111/maps.12895.} \end{APACrefDOI}
\PrintBackRefs{\CurrentBib}

\bibitem [\protect \citeauthoryear {%
Robbins%
\ \BBA {} Hynek%
}{%
Robbins%
\ \BBA {} Hynek%
}{%
{\protect \APACyear {2012}}%
}]{%
Robbins2012a}
\APACinsertmetastar {%
Robbins2012a}%
\begin{APACrefauthors}%
Robbins, S\BPBI J.%
\BCBT {}\ \BBA {} Hynek, B\BPBI M.%
\end{APACrefauthors}%
\unskip\
\newblock
\APACrefYearMonthDay{2012}{}{}.
\newblock
{\BBOQ}\APACrefatitle {A New Global Database of {{Mars}} Impact Craters Larger
  than 1 Km: 2. {{Global}} Crater Properties and Regional Variations of the
  Simple-to-Complex Transition Diameter} {A new global database of {{Mars}}
  impact craters larger than 1 km: 2. {{Global}} crater properties and regional
  variations of the simple-to-complex transition diameter}.{\BBCQ}
\newblock
\APACjournalVolNumPages{Journal of Geophysical Research E:
  Planets}{117}{6}{1--21}.
\newblock
\begin{APACrefDOI} \doi{10.1029/2011JE003967} \end{APACrefDOI}
\PrintBackRefs{\CurrentBib}

\bibitem [\protect \citeauthoryear {%
Salamuni{\'c}car%
, Lon{\v c}ari{\'c}%
\BCBL {}\ \BBA {} Mazarico%
}{%
Salamuni{\'c}car%
\ \protect \BOthers {.}}{%
{\protect \APACyear {2012}}%
}]{%
Salamuniccar2012}
\APACinsertmetastar {%
Salamuniccar2012}%
\begin{APACrefauthors}%
Salamuni{\'c}car, G.%
, Lon{\v c}ari{\'c}, S.%
\BCBL {}\ \BBA {} Mazarico, E.%
\end{APACrefauthors}%
\unskip\
\newblock
\APACrefYearMonthDay{2012}{}{}.
\newblock
{\BBOQ}\APACrefatitle {{{LU60645GT}} and {{MA132843GT}} Catalogues of {{Lunar}}
  and {{Martian}} Impact Craters Developed Using a {{Crater Shape-based}}
  Interpolation Crater Detection Algorithm for Topography Data} {{{LU60645GT}}
  and {{MA132843GT}} catalogues of {{Lunar}} and {{Martian}} impact craters
  developed using a {{Crater Shape-based}} interpolation crater detection
  algorithm for topography data}.{\BBCQ}
\newblock
\APACjournalVolNumPages{Planetary and Space Science}{60}{1}{236--247}.
\newblock
\begin{APACrefDOI} \doi{10.1016/j.pss.2011.09.003} \end{APACrefDOI}
\PrintBackRefs{\CurrentBib}

\bibitem [\protect \citeauthoryear {%
Silburt%
\ \protect \BOthers {.}}{%
Silburt%
\ \protect \BOthers {.}}{%
{\protect \APACyear {2019}}%
}]{%
Silburt2019}
\APACinsertmetastar {%
Silburt2019}%
\begin{APACrefauthors}%
Silburt, A.%
, {Ali-Dib}, M.%
, Zhu, C.%
, Jackson, A.%
, Valencia, D.%
, Kissin, Y.%
\BDBL {}Menou, K.%
\end{APACrefauthors}%
\unskip\
\newblock
\APACrefYearMonthDay{2019}{}{}.
\newblock
{\BBOQ}\APACrefatitle {Lunar Crater Identification via Deep Learning} {Lunar
  crater identification via deep learning}.{\BBCQ}
\newblock
\APACjournalVolNumPages{Icarus}{317}{}{27--38}.
\newblock
\begin{APACrefDOI} \doi{10.1016/j.icarus.2018.06.022} \end{APACrefDOI}
\PrintBackRefs{\CurrentBib}

\bibitem [\protect \citeauthoryear {%
Yang%
\ \BBA {} Cai%
}{%
Yang%
\ \BBA {} Cai%
}{%
{\protect \APACyear {2022}}%
}]{%
yangHighResolutionFeaturePyramid2022}
\APACinsertmetastar {%
yangHighResolutionFeaturePyramid2022}%
\begin{APACrefauthors}%
Yang, S.%
\BCBT {}\ \BBA {} Cai, Z.%
\end{APACrefauthors}%
\unskip\
\newblock
\APACrefYearMonthDay{2022}{}{}.
\newblock
{\BBOQ}\APACrefatitle {High-{{Resolution Feature Pyramid Network}} for
  {{Automatic Crater Detection}} on {{Mars}}} {High-{{Resolution Feature
  Pyramid Network}} for {{Automatic Crater Detection}} on {{Mars}}}.{\BBCQ}
\newblock
\APACjournalVolNumPages{IEEE Transactions on Geoscience and Remote
  Sensing}{60}{}{1--12}.
\newblock
\begin{APACrefDOI} \doi{10.1109/TGRS.2021.3104925} \end{APACrefDOI}
\PrintBackRefs{\CurrentBib}

\end{thebibliography}

\end{document}